\documentclass[prb,twocolumn,showpacs,superscriptaddress,floatfix,amsfonts,amsmath]{revtex4}

\usepackage{graphicx,subfigure}

\newcommand{\be}{\begin{equation}}
\newcommand{\ee}{\end{equation}}
\newcommand{\bea}{\begin{eqnarray}}
\newcommand{\eea}{\end{eqnarray}}
\newcommand{\down}{\downarrow}

\newcommand{\cobaltchloride}{\rm{CoCl}_2\cdot 2\rm{H}_2\rm{O}}
\newcommand{\cobaltniobate}{\rm{CoNb}_2\rm{O}_6}

\begin{document}

\title{Numerical Simulations of Laser Induced Magnetic Bloch Oscillations}
\author{Sergey \surname{Shinkevich}}
\author{Olav  F. \surname{Sylju{\aa}sen}}
\affiliation{Department of Physics, University of Oslo, P.~O.~Box 1048 Blindern, N-0316 Oslo, Norway}

\date{\today}

\pacs{75.78.-n,75.40.Mg,75.10.Pq,78.47.D-}

\begin{abstract}
We propose to use a laser to generate magnetic Bloch oscillations in one dimensional easy-axis ferromagnets at low temperatures. This proposal is investigated numerically in details for material parameters relevant for $\rm{CoCl}_2\cdot 2\rm{H}_2\rm{O}$.
\end{abstract}

\maketitle

According to quantum mechanics a particle in a periodic potential will oscillate in response to a constant force. Such Bloch oscillations (BO) were predicted in the early days of quantum mechanics\cite{Bloch,Zener}, but have only recently been experimentally demonstrated in very clean semiconductor superlattices\cite{BOsemiconductors} and in Bose-Einstein condensates\cite{BOcondensates}. 

In condensed matter systems the particles need not be of the ordinary kind resembling electrons. In particular, an elementary particle in a one dimensional anisotropic ferromagnet is a domain-wall separating regions of up and down spins. Such domain-walls can have a dispersion relation like that derived from a periodic potential, and in the presence of a uniform magnetic field these magnetic systems have been predicted to show BO\cite{KyriakidisLoss}.
In particular the blue crystalline material $\cobaltchloride$ and $\cobaltniobate$ have been proposed as candidate materials. However, no BO have been experimentally observed in these to date.

In a recent article\cite{Shinkevich} we have revisited the material $\cobaltchloride$ and refined the predictions of Ref.~\onlinecite{KyriakidisLoss} for detecting spectral signatures of BO in neutron scattering experiments, taking also into account extra interactions present in the material. While our result indicates that the spectral signatures of BO can indeed be observed in neutron scattering, the signatures are relatively weak, at the 10\% level of the total spectral weight at finite temperatures. This is a consequence of neutrons being a relatively weak probe as they cause only single spin-flip excitations.

We propose here a more direct way to generate BO by keeping the material at low temperature and induce excitations using a short laser-pulse. Upon turning off the laser-pulse the magnetization of the material will continue oscillating at the Bloch frequency. 
It has been known since long ago\cite{TorranceTinkham,NicoliTinkham} that light in the far-infrared frequency range can induce magnetic excitations in $\cobaltchloride$, but no time-dependence of the magnetization was studied there.
 
 In this article we model the laser-pulse as a time-dependent perturbation to the Hamiltonian and investigate its effects by solving the time-dependent Schr{\"o}dinger equation numerically. We show that BO can be generated this way, and give appropriate laser frequencies and pulse-duration times. 
  
The magnetic properties of $\cobaltchloride$ are described by the spin-1/2 Hamiltonian
\be
    H = - \sum_{i} \left( J_z S^z_i S^z_{i+1} + h_z S^z_i \right) + H_d,
\ee
where $H_d$ denotes subdominant terms to be discussed below.
The ferromagnetic coupling $J_z=36.5\, {\rm K}$\cite{TorranceTinkham} is the dominant term in the Hamiltonian. Alone it causes neighboring spins to align their spin z-components, thus the energy of an excited state depends on the number of anti-aligned spin neighbors;  domain-walls, where each domain-wall costs an energy of $J_z/2$. In the presence of an external magnetic field along the z-axis, $h_z$, the energy will also depend on the number of spins opposing the field, implying pair-wise confinement of domain-walls. Such a bound-state of two domain-walls separating $l$ overturned spins, known as a spin cluster excitation\cite{TorranceTinkham}, or simply a domain, has an energy $J_z + h_z l$ above the ground state energy.

The term $H_d$ describes additional couplings that partly give dynamics to the domain walls and partly induce more domain-walls,
\bea
H_d & = & -\sum_{i}\left[ J_a \left( S^+_i S^+_{i+1} + S^-_i S^-_{i+1} \right) \right.  \nonumber \\
          &   & \left. \qquad + J_\perp \left( S^+_i S^-_{i+1} + S^-_i S^+_{i+1} \right) \right],
\eea
where $J_a=3.8\,{\rm K}$\cite{Montfrooij} and $J_\perp=5.43\,{\rm K}$\cite{Montfrooij}. $S^{\pm}_i= S^x_i \pm i S^y_i$ are the usual raising and lowering operators. The $J_a$-term can move a domain-wall two lattice spacings, thus mixing states with even (or odd) $l$. Similarly $J_\perp$ gives kinetic energy to the $l=1$ domain state. Both of these terms can also induce new domain-wall pairs. However, with $J_z$ being the dominant coupling, extra domain-walls will be energetically costly, thus we will restrict our calculations to states having a small number of domain-walls $N_{dw}$.

When restricting to $N_{dw} \le 2$, $H$ can be diagonalized. The energy spectrum is $E_n = J_z + \mu_n h_z$, where $\mu_n$ is found by solving an equation involving a ratio of Bessel functions\cite{Shinkevich}. For high energies $\mu_n \approx n$, an integer, thus the energy spectrum becomes equidistant. The corresponding energy eigenfunctions $|\psi_n \rangle$ are Bessel functions\cite{Shinkevich}. 
From these one can construct a time-dependent state $ | \chi(t) \rangle = \sum_n a_n e^{-iE_n t} |\psi_n \rangle$. When this state is dominated by energy eigenstates with energies from the equidistant region, the time-dependence of the magnetization $M_z(t) = \langle \chi(t) |\sum_i S^z_i |\chi(t) \rangle$  becomes
\be \label{besseloscillations}
   M_z(t) = c- 2 x_0 \sum_n Re\left( a^*_n a_{n+2} e^{-i \omega_B t} \right)
 \ee
where $c$ is a constant, $\omega_B=2h_z$ is the Bloch frequency, and $x_0=2J_a/h_z$. Thus the BO amplitude is proportional to $4J_a/h_z$ times a factor which depends on the probability amplitudes $a_n$ of the excited states. Our BO amplitude is a factor 2 larger than obtained in Ref.~\onlinecite{KyriakidisLoss} as we consider the size oscillations of a single domain having two domain-walls, while they considered the motion of a single domain-wall. 

\begin{figure}
\includegraphics[clip,width=8cm]{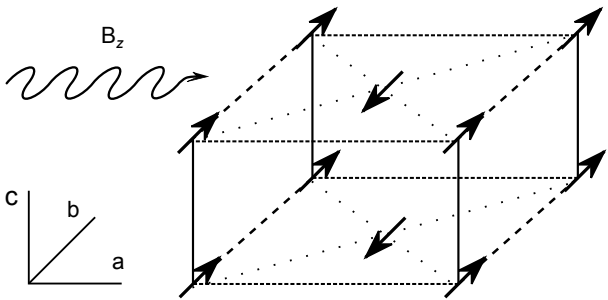} 
\caption{In $\cobaltchloride$ the spin z-axis corresponds to the crystallographic b axis. The strong ferromagnetic $J_z$ couples spins along the c axis. A laser beam is shown propagating at normal incidence to the b-c-plane with magnetic field polarization in the b direction.}
\label{orientation}
\end{figure}

In order to populate the excited levels at very low temperatures, thus producing BO, we propose here to use a laser with a wavelength in the far-infrared. The ${\rm Co}$ electrons causing the magnetism are d-shell electrons thus having no electric dipole moment. We therefore model the laser as an extra time-dependent magnetic field which couples to the spins as
\be
    H_{ext} = -B^z_0 \cos( \omega t) \sum_i S^z_i,
\ee
where $\omega$ is the laser frequency and $B^z_0$ is the laser magnetic field amplitude. We have here assumed a linearly polarized laser beam such that the magnetic field is along the Ising direction. This corresponds to the crystallographic $b$ direction in $\cobaltchloride$. Such a setup can be made by cleaving the crystals in the b-c plane and directing the laser at normal incidence to this surface polarizing the laser beam such that the magnetic field points along the b-direction, see Fig.~\ref{orientation}. We have assumed the laser beam to be coherent along its front, and also through the crystal. The laser considered here have a wavelength of about $0.3\,{\rm mm}$ thus for this approximation to be good the crystal should be thinner than this. For thicker crystals there will be an additional phase-shift associated with the depth. 

The time-dependent perturbation can be treated numerically for large system sizes when restricting to states where $N_{dw} \leq 2$.  
This restriction implies that the energy gap between the ground state and any excited state will depend on the system size $N$. This can be understood by considering the perturbative energy correction from virtual processes involving the creation and destruction of an additional domain. As there are roughly $N$ places to insert the new domain, the energy correction will be proportional to $N$. When restricting to $N_{dw} \leq 2$ the ferromagnetic state receives this correction, but not the states having one domain, as their corrections come from the excluded $N_{dw} \geq 4$ sector. In order to make the energy gap intensive we redefine the coupling between the ferromagnetic and the $l=2$, $N_{dw}=2$ state in the Hamiltonian by dividing it by a factor $\sqrt{N}$.\cite{Shinkevich} This effectively makes the correction to the ferromagnetic state independent of system size. 

\begin{figure}
\includegraphics[clip,width=8cm]{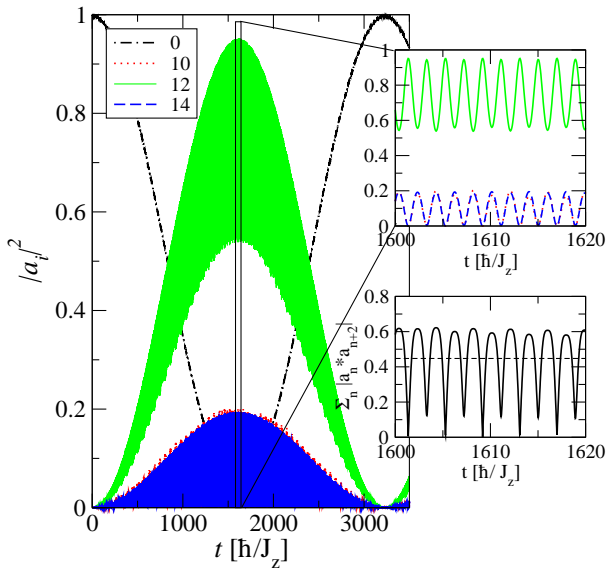}
\caption{(color online) Population $|a_j|^2$ of selected energy levels $j$, indicated in the legend, as a function of time after turning on the laser, $H_{ext}$. $\omega=(E_{12}-E_0)/\hbar$ and $B_0^z=0.2J_z$. The upper right panel shows a zoom in on the boxed time region. The lower right panel shows the time dependence of the sum of products of two nearby probability amplitudes, $\sum_n |a_n^* a_{n+2}|$. The time averaged value is shown as the horizontal dashed line. $t$ is measured in units of $\hbar/J_z$ which is $0.2 \cdot 10^{-12} {\rm s}$ for $\cobaltchloride$.}
\label{population}
\end{figure}

Starting in the ground state of $H$, the time-dependent Schr{\"o}dinger equation is solved iteratively numerically with the laser field $H_{ext}$ present. In the iterations we keep the 300 lowest energy states of $H$ with zero momentum and $N_{dw} \leq 2$. The laser frequency $\omega$ is tuned such that $\omega=E_n-E_0$ where $n$ corresponds to an energy level in the region where the spectrum is approximately equidistant. We choose $n=12$ corresponding to $E_{12}-E_0 \approx 1.6J_z$ for a static magnetic field $h_z=0.05J_z$. In practice when using a laser with a fixed wavelength, resonance can instead be found by changing the static magnetic field thereby adjusting the energy levels. The iterative solution gives time-dependent amplitudes of the different energy levels. Fig.~\ref{population} shows the probabilities of finding the system in selected levels as a function of time. Only even $n$ states are excited because the $S^z$-terms do not flip any spins.
The black dot-dashed curve in Fig.~\ref{population} shows how the ground state is depleted. The minimum of the ground state population coincides with the maximum of the population of level $n=12$, green solid curve, and occurs at a time $\tau=\pi \hbar/\omega_R$ where $\omega_R=\sqrt{|B^z_0 \alpha_{12}|^2+(\omega-(E_{12}-E_0))^2}$, where $\alpha_{12}$ is the matrix element of $\sum_i S^z_i$ between the ground state and the $n=12$ excited state. These oscillations are in essence Rabi oscillations. 
Exciting the level $n=12$ alone does not give appreciable amplitude for Bloch oscillations as one also needs to populate the levels with $n+2$ (or $n-2$), see Eq.~\ref{besseloscillations}. This can be achieved by using a large amplitude of the laser, we used $B^z_0=0.2J_z$, thereby causing off-resonant tunneling between the $n$ and the $n \pm 2$ levels, see Fig.~\ref{population}, red dotted and blue dashed curves. These off-resonant tunneling processes are fast, thus the population of the nearby levels follows closely that of the central level. 

In order to seek the maximum amplitude of BO we turn off $H_{ext}$ at the maximum population of the central excited level. A close look at the time dependence of  $a^*_n a_{n+2}$ near the cutoff time, reveals that the dominating terms have the same phase, thus the amplitude of BO is proportional to $\sum_n |a_n^* a_{n+2}|$ which is shown in the lower inset of Fig.~\ref{population}. From this we see that it oscillates fast with a frequency corresponding to $E_{12}-E_0$. It may be difficult to turn off the laser when this quantitiy is maximal. However, this is not a major concern as the time averaged value is about $75\%$ of the maximum value. 

\begin{figure}
\includegraphics[clip,width=8cm]{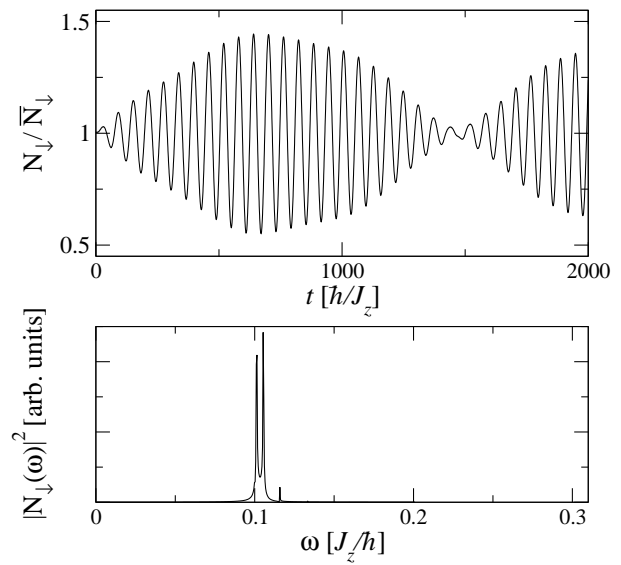}
\caption{Size oscillations of a domain excitation, measured as the number of down spins $N_{1\down}/\bar{N}_{1\down}$ vs. time after the laser is switched off (upper panel), and its Fourier spectrum (lower panel).}
\label{magnetization}
\end{figure}

Turning off the laser at a time $\tau$ when $\sum_n |a_n^* a_{n+2}|$ is maximal, and letting the system evolve further in time without $H_{ext}$, produces the BO shown in Fig.~\ref{magnetization}. As our simulation only allows single domain excitations we have plotted the relative size of the domain, measured by the expectation value of the number of spins opposing the field $N_{1\down}(t)= M_z(t)-N/2$ divided by its time average $\bar{N}_{1\down} \approx n$ for excitation $E_n$.  We see that the relative size of the domain oscillates between $0.6$ and $1.4$ corresponding to a size between $7.2$ and $16.8$ for $\bar{N}_{12}=12$. Thus the amplitude is $4.8$ which is close to the expected value $4 J_a/h_z \times 0.6=5$ from Eq.~\ref{besseloscillations}. Allowing a finite density $\rho$ of coherently oscillating domain states, the relative size of a single domain state shown in Fig.~\ref{magnetization} will be proportional to the experimentally relevant quantity,  the time-dependent relative magnetization: $(M_z(t) -\bar{M})/ \bar{M} = (\rho {\bar N}_{1\down}/(1/2-\rho \bar{N}_{1\down}) (N_{1\down}(t)/\bar{N}_{1\down}-1 )$. 

The BO in Fig.~{\ref{magnetization} are not simple harmonic. A Fourier transformation of the beating pattern is shown in the lower panel of Fig.~\ref{magnetization}. Two peaks are clearly visible. They correspond to $\omega_1 = E_{14}-E_{12}= 0.1J_z$ and $\omega_2 = E_{12}-E_{10} = 0.11 J_z$. Thus the beating pattern is due to the deviation from an equidistant ladder spectrum. This frequency difference can be made smaller by exciting higher energy bound states where the spectrum is closer to being equidistant.

In this setup we rely on off-resonant tunneling in order to populate nearby levels. This requires a large laser amplitude. Instead one might use two small-amplitude lasers each in resonance with nearby levels.  One can also change the laser polarization to have a component along the spin-$x$ direction. This will induce transitions between the even and odd $n$ states. Our simulations show BO also in this case, but now with more frequency components due to the deviation from equidistant spectra, as in this case both the even and the odd sectors participate.   

The restriction to $N_{dw} \leq 2$ and the associated redefinition of the coupling to the ferromangetic state can raise doubts about the validity of the matrix elements calculation, also it does not allow for any discussion of interactions between domains. 
We have therefore numerically also investigated cases where we allow more domains without any redefinition of couplings. Computer performance restrictions let us consider $N_{dw} \leq 6$ for $N\leq 34$. In Fig.~\ref{energy} it is seen that the $N$ dependence of the energy gap $\Delta$ to the first excited state decreases as higher domain wall sectors are included. For comparison the $N_{dw} \leq 2$ result with redefined coupling to the ferromagnetic state is shown as the dashed line.
\begin{figure}
\includegraphics[clip,width=8cm]{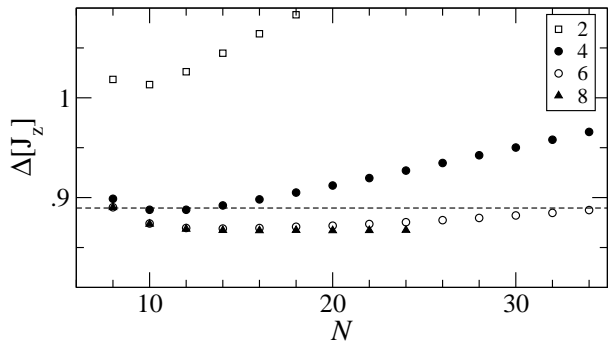} 
\caption{Energy gap $\Delta$ to the first excited state vs. system size $N$ for different $N_{dw}$. The legend specifies the maximum $N_{dw}$ included in the diagonalization. The dashed line shows the energy gap for $N_{dw} \leq 2$ with the $1/\sqrt{N}$ redefinition of the coupling to the ferromagnetic state.}
\label{energy}
\end{figure}

In order to identify the elementary domain excitations and the associated transition matrix element of the laser field between these and the ground state, we construct the approximate single domain creation operator\cite{Shinkevich} 
\be
    a_{p,n}^\dagger = \sum_{l,j} e^{-ipl/2} J_{(l-\mu_n)/2}\left( x_0 \cos p \right) \Pi_{k=j}^{j+l-1} S^+_{k}
\ee
with momentum $p=0$ and let it act on the ferromagnetic state. The sum is restricted to even values of $l$ and $J_{m}$ is the Bessel function of the first kind of order $m$. We then compute the overlap of this with the exact eigenstates of the $N_{dw} \leq 6$ system. 
For each value of $n$ we pick the state with maximum overlap. Fig.~\ref{quasiparticles}a) shows the energies of these states. We see that for high enough $n$ the energies become equidistant and agree well with what we found for $N_{dw} \leq 2$. The laser transition matrix element between these states and the ground state behaves as $\sqrt{N} \alpha_n$. The coefficients $\alpha_n$ are very close to those we found for $N_{dw} \leq 2$, see Fig.~\ref{quasiparticles}a), thus the Rabi frequency is not changed provided interactions between domains are negligible. Note that $\alpha_n$ drops very fast with increasing $n$. 
\begin{figure}
\includegraphics[clip,width=8cm]{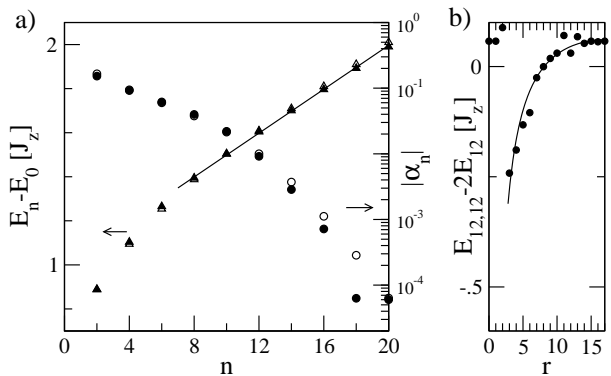} 
\caption{a) Excitation energy of elementary domain states vs. $n$ (triangles). Also shown is $|\alpha_n|$ vs. $n$ (circles) on a semi-log scale. Filled symbols refer to $N_{dw} \leq 6$ and $N=34$, while open symbols are for $N_{dw} \leq 2$ with redefined coupling to the ferromagnetic state.   
b) Interaction energy of two $n=12$ excitations vs. separation distance $r$ for $N_{dw} \leq 6$ and $N=34$.}   
\label{quasiparticles}
\end{figure}

Insights about interactions can be gotten by identifying two-domain excitations. We construct approximate two-domain states as pairs with total momentum zero of two $n=12$ single domain states separated by a distance $r$, with creation operator $b^\dagger_{r,n} = \sum_{p} a^\dagger_{p,n} a^\dagger_{-p,n}  e^{i pr}$. We let this act on the ferromagnetic state, retain only the terms having four domain-walls, and compute its overlap with the exact eigenstates of the $N_{dw} \leq 6$ system. The energy of the states with maximum overlap minus two times the single-particle excitation energy $E_{12}$ are shown in Fig.~\ref{quasiparticles}b) as a function of $r$. We interpret this as the interaction energy of domains separated by $r$ lattice spacings. We find that the functional form $a-b[1/r + 1/(N-r)]$ fits the results reasonably with $a=0.22J_z$ and $b=1.37J_z$. The positive $a$ is caused by the restriction on $N_{dw} $ which tends to overestimate energies in higher domain-wall sectors relative to those in lower sectors. 
This interaction energy causes an inhomogeneous broadening of the resonance frequency, and can lead to an upper limit on the density of domains excited by the laser together with an increased Rabi frequency  due to interaction blocking effects\cite{Lukin,Rydberg}.  It can also change the BO frequency locally. However, in order to study this one needs to look at {\em differences} between interaction energies for neighboring energy levels.  We have not been able to study long enough chains to address this issue with sufficient precision.  The lifetime of a domain excitation can also be reduced due to collisions, however the domain excitations are heavy due to their flat dispersion, so we expect a significant lifetime reduction only at a high density of excitations when neighboring domains are touching. 

We conclude that it should be possible to excite magnetic BO in $\cobaltchloride$ using a laser at low temperatures in a static magnetic field. In addition our simulations using the Hamiltonian and parameters from Ref.~\onlinecite{Kjall} also indicate that BO in $\cobaltniobate$ may be generated in a similar way.     

\end{document}